\documentclass[twocolumn,letterpaper,aps,prd,longbibliography,superscriptaddress,floatfix]{revtex4-1}

\usepackage{graphicx}   
\usepackage{xspace}     
\usepackage{epstopdf}

\begin{document}


\title{Measurement of charged pion double spin asymmetries at 
midrapidity in longitudinally polarized $p$$+$$p$ collisions 
at $\sqrt{s}=510$~GeV}

\newcommand{\abilene}{Abilene Christian University, Abilene, Texas 79699, USA}
\newcommand{\augie}{Department of Physics, Augustana University, Sioux Falls, South Dakota 57197, USA}
\newcommand{\banaras}{Department of Physics, Banaras Hindu University, Varanasi 221005, India}
\newcommand{\barc}{Bhabha Atomic Research Centre, Bombay 400 085, India}
\newcommand{\baruch}{Baruch College, City University of New York, New York, New York, 10010 USA}
\newcommand{\bnlcoll}{Collider-Accelerator Department, Brookhaven National Laboratory, Upton, New York 11973-5000, USA}
\newcommand{\bnlphys}{Physics Department, Brookhaven National Laboratory, Upton, New York 11973-5000, USA}
\newcommand{\caucr}{University of California-Riverside, Riverside, California 92521, USA}
\newcommand{\charlesczech}{Charles University, Ovocn\'{y} trh 5, Praha 1, 116 36, Prague, Czech Republic}
\newcommand{\cns}{Center for Nuclear Study, Graduate School of Science, University of Tokyo, 7-3-1 Hongo, Bunkyo, Tokyo 113-0033, Japan}
\newcommand{\colorado}{University of Colorado, Boulder, Colorado 80309, USA}
\newcommand{\columbia}{Columbia University, New York, New York 10027 and Nevis Laboratories, Irvington, New York 10533, USA}
\newcommand{\czechtech}{Czech Technical University, Zikova 4, 166 36 Prague 6, Czech Republic}
\newcommand{\debrecen}{Debrecen University, H-4010 Debrecen, Egyetem t{\'e}r 1, Hungary}
\newcommand{\elte}{ELTE, E{\"o}tv{\"o}s Lor{\'a}nd University, H-1117 Budapest, P{\'a}zm{\'a}ny P.~s.~1/A, Hungary}
\newcommand{\eszterhazy}{Eszterh\'azy K\'aroly University, K\'aroly R\'obert Campus, H-3200 Gy\"ongy\"os, M\'atrai \'ut 36, Hungary}
\newcommand{\ewha}{Ewha Womans University, Seoul 120-750, Korea}
\newcommand{\famu}{Florida A\&M University, Tallahassee, FL 32307, USA}
\newcommand{\fsu}{Florida State University, Tallahassee, Florida 32306, USA}
\newcommand{\gsu}{Georgia State University, Atlanta, Georgia 30303, USA}
\newcommand{\hanyang}{Hanyang University, Seoul 133-792, Korea}
\newcommand{\hiroshima}{Hiroshima University, Kagamiyama, Higashi-Hiroshima 739-8526, Japan}
\newcommand{\howard}{Department of Physics and Astronomy, Howard University, Washington, DC 20059, USA}
\newcommand{\ihepprot}{IHEP Protvino, State Research Center of Russian Federation, Institute for High Energy Physics, Protvino, 142281, Russia}
\newcommand{\illuiuc}{University of Illinois at Urbana-Champaign, Urbana, Illinois 61801, USA}
\newcommand{\inrras}{Institute for Nuclear Research of the Russian Academy of Sciences, prospekt 60-letiya Oktyabrya 7a, Moscow 117312, Russia}
\newcommand{\instpasczech}{Institute of Physics, Academy of Sciences of the Czech Republic, Na Slovance 2, 182 21 Prague 8, Czech Republic}
\newcommand{\isu}{Iowa State University, Ames, Iowa 50011, USA}
\newcommand{\jaea}{Advanced Science Research Center, Japan Atomic Energy Agency, 2-4 Shirakata Shirane, Tokai-mura, Naka-gun, Ibaraki-ken 319-1195, Japan}
\newcommand{\jeonbuk}{Jeonbuk National University, Jeonju, 54896, Korea}
\newcommand{\jyvaskyla}{Helsinki Institute of Physics and University of Jyv{\"a}skyl{\"a}, P.O.Box 35, FI-40014 Jyv{\"a}skyl{\"a}, Finland}
\newcommand{\kek}{KEK, High Energy Accelerator Research Organization, Tsukuba, Ibaraki 305-0801, Japan}
\newcommand{\korea}{Korea University, Seoul 02841, Korea}
\newcommand{\kurchatov}{National Research Center ``Kurchatov Institute", Moscow, 123098 Russia}
\newcommand{\kyoto}{Kyoto University, Kyoto 606-8502, Japan}
\newcommand{\lahorelums}{Physics Department, Lahore University of Management Sciences, Lahore 54792, Pakistan}
\newcommand{\lawllnl}{Lawrence Livermore National Laboratory, Livermore, California 94550, USA}
\newcommand{\losalamos}{Los Alamos National Laboratory, Los Alamos, New Mexico 87545, USA}
\newcommand{\lund}{Department of Physics, Lund University, Box 118, SE-221 00 Lund, Sweden}
\newcommand{\lyon}{IPNL, CNRS/IN2P3, Univ Lyon, Universit?? Lyon 1, F-69622, Villeurbanne, France}
\newcommand{\maryland}{University of Maryland, College Park, Maryland 20742, USA}
\newcommand{\mass}{Department of Physics, University of Massachusetts, Amherst, Massachusetts 01003-9337, USA}
\newcommand{\michigan}{Department of Physics, University of Michigan, Ann Arbor, Michigan 48109-1040, USA}
\newcommand{\muhlenberg}{Muhlenberg College, Allentown, Pennsylvania 18104-5586, USA}
\newcommand{\myongji}{Myongji University, Yongin, Kyonggido 449-728, Korea}
\newcommand{\nara}{Nara Women's University, Kita-uoya Nishi-machi Nara 630-8506, Japan}
\newcommand{\natmephi}{National Research Nuclear University, MEPhI, Moscow Engineering Physics Institute, Moscow, 115409, Russia}
\newcommand{\newmex}{University of New Mexico, Albuquerque, New Mexico 87131, USA}
\newcommand{\nmsu}{New Mexico State University, Las Cruces, New Mexico 88003, USA}
\newcommand{\northcg}{Physics and Astronomy Department, University of North Carolina at Greensboro, Greensboro, North Carolina 27412, USA}
\newcommand{\ohio}{Department of Physics and Astronomy, Ohio University, Athens, Ohio 45701, USA}
\newcommand{\ornl}{Oak Ridge National Laboratory, Oak Ridge, Tennessee 37831, USA}
\newcommand{\orsay}{IPN-Orsay, Univ.~Paris-Sud, CNRS/IN2P3, Universit\'e Paris-Saclay, BP1, F-91406, Orsay, France}
\newcommand{\peking}{Peking University, Beijing 100871, People's Republic of China}
\newcommand{\pnpi}{PNPI, Petersburg Nuclear Physics Institute, Gatchina, Leningrad region, 188300, Russia}
\newcommand{\pusan}{Pusan National University, Busan, 46241, South Korea}
\newcommand{\riken}{RIKEN Nishina Center for Accelerator-Based Science, Wako, Saitama 351-0198, Japan}
\newcommand{\rikjrbrc}{RIKEN BNL Research Center, Brookhaven National Laboratory, Upton, New York 11973-5000, USA}
\newcommand{\rikkyo}{Physics Department, Rikkyo University, 3-34-1 Nishi-Ikebukuro, Toshima, Tokyo 171-8501, Japan}
\newcommand{\saispbstu}{Saint Petersburg State Polytechnic University, St.~Petersburg, 195251 Russia}
\newcommand{\seoulnat}{Department of Physics and Astronomy, Seoul National University, Seoul 151-742, Korea}
\newcommand{\stonybrkc}{Chemistry Department, Stony Brook University, SUNY, Stony Brook, New York 11794-3400, USA}
\newcommand{\stonycrkp}{Department of Physics and Astronomy, Stony Brook University, SUNY, Stony Brook, New York 11794-3800, USA}
\newcommand{\tenn}{University of Tennessee, Knoxville, Tennessee 37996, USA}
\newcommand{\titech}{Department of Physics, Tokyo Institute of Technology, Oh-okayama, Meguro, Tokyo 152-8551, Japan}
\newcommand{\tsukuba}{Tomonaga Center for the History of the Universe, University of Tsukuba, Tsukuba, Ibaraki 305, Japan}
\newcommand{\vandy}{Vanderbilt University, Nashville, Tennessee 37235, USA}
\newcommand{\weizmann}{Weizmann Institute, Rehovot 76100, Israel}
\newcommand{\wigner}{Institute for Particle and Nuclear Physics, Wigner Research Centre for Physics, Hungarian Academy of Sciences (Wigner RCP, RMKI) H-1525 Budapest 114, POBox 49, Budapest, Hungary}
\newcommand{\yonsei}{Yonsei University, IPAP, Seoul 120-749, Korea}
\newcommand{\zagreb}{Department of Physics, Faculty of Science, University of Zagreb, Bijeni\v{c}ka c.~32 HR-10002 Zagreb, Croatia}
\affiliation{\abilene}
\affiliation{\augie}
\affiliation{\banaras}
\affiliation{\barc}
\affiliation{\baruch}
\affiliation{\bnlcoll}
\affiliation{\bnlphys}
\affiliation{\caucr}
\affiliation{\charlesczech}
\affiliation{\cns}
\affiliation{\colorado}
\affiliation{\columbia}
\affiliation{\czechtech}
\affiliation{\debrecen}
\affiliation{\elte}
\affiliation{\eszterhazy}
\affiliation{\ewha}
\affiliation{\famu}
\affiliation{\fsu}
\affiliation{\gsu}
\affiliation{\hanyang}
\affiliation{\hiroshima}
\affiliation{\howard}
\affiliation{\ihepprot}
\affiliation{\illuiuc}
\affiliation{\inrras}
\affiliation{\instpasczech}
\affiliation{\isu}
\affiliation{\jaea}
\affiliation{\jeonbuk}
\affiliation{\jyvaskyla}
\affiliation{\kek}
\affiliation{\korea}
\affiliation{\kurchatov}
\affiliation{\kyoto}
\affiliation{\lahorelums}
\affiliation{\lawllnl}
\affiliation{\losalamos}
\affiliation{\lund}
\affiliation{\lyon}
\affiliation{\maryland}
\affiliation{\mass}
\affiliation{\michigan}
\affiliation{\muhlenberg}
\affiliation{\myongji}
\affiliation{\nara}
\affiliation{\natmephi}
\affiliation{\newmex}
\affiliation{\nmsu}
\affiliation{\northcg}
\affiliation{\ohio}
\affiliation{\ornl}
\affiliation{\orsay}
\affiliation{\peking}
\affiliation{\pnpi}
\affiliation{\pusan}
\affiliation{\riken}
\affiliation{\rikjrbrc}
\affiliation{\rikkyo}
\affiliation{\saispbstu}
\affiliation{\seoulnat}
\affiliation{\stonybrkc}
\affiliation{\stonycrkp}
\affiliation{\tenn}
\affiliation{\titech}
\affiliation{\tsukuba}
\affiliation{\vandy}
\affiliation{\weizmann}
\affiliation{\wigner}
\affiliation{\yonsei}
\affiliation{\zagreb}
\author{U.~Acharya} \affiliation{\gsu} 
\author{A.~Adare} \affiliation{\colorado} 
\author{C.~Aidala} \affiliation{\michigan} 
\author{N.N.~Ajitanand} \altaffiliation{Deceased} \affiliation{\stonybrkc} 
\author{Y.~Akiba} \email[PHENIX Spokesperson: ]{akiba@rcf.rhic.bnl.gov} \affiliation{\riken} \affiliation{\rikjrbrc} 
\author{R.~Akimoto} \affiliation{\cns} 
\author{M.~Alfred} \affiliation{\howard} 
\author{N.~Apadula} \affiliation{\isu} \affiliation{\stonycrkp} 
\author{Y.~Aramaki} \affiliation{\riken} 
\author{H.~Asano} \affiliation{\kyoto} \affiliation{\riken} 
% ...
\collaboration{PHENIX Collaboration} \noaffiliation

\date{\today}


\begin{abstract}


The PHENIX experiment at the Relativistic Heavy Ion Collider has 
measured the longitudinal double spin asymmetries, $A_{LL}$, for charged 
pions at midrapidity ($|\eta|<$~0.35) in longitudinally polarized 
$p$$+$$p$ collisions at $\sqrt{s}=510$~GeV. These measurements are 
sensitive to the gluon spin contribution to the total spin of the proton 
in the parton momentum fraction $x$ range between 0.04 and 0.09. One can 
infer the sign of the gluon polarization from the ordering of pion 
asymmetries with charge alone. The asymmetries are found to be 
consistent with global quantum-chromodynamics fits of deep-inelastic 
scattering and data at $\sqrt{s}=200$~GeV, which show a nonzero 
positive contribution of gluon spin to the proton spin.

\end{abstract}

\maketitle

\section{Introduction}

The spin of the proton is known to be $\hbar/2$, yet its decomposition 
in terms of its constituents, quarks and gluons, is not very well known. 
Initially, the fixed-target deep-inelastic scattering (DIS) experiments 
measured the polarized structure function, $g_1(x,Q^2)$, where $x$ is 
the parton momentum fraction of the proton and $Q^{2}$ is the momentum 
transfer squared, enabling the reconstruction of the quark spin 
contributions, $\Delta \Sigma(x,Q^2)$, with the help of weak and hyperon 
decay constants. Early measurements found this contribution to be 
substantially smaller than expected~\cite{Ashman:1987hv}, leading to the 
so-called spin crisis. In addition to the quark spins, gluon spins as 
well as the constituents' orbital angular momenta can contribute to the 
spin sum rule~\cite{Jaffe:1989jz}. Because DIS at low to moderate energies 
essentially couples through the electromagnetic interaction, it is most 
sensitive to the quark spin contributions and the gluon spin only enters 
via scaling violations.

In contrast, in polarized $p$$+$$p$ collisions, for example at the 
Relativistic Heavy Ion Collider (RHIC), the dominant hard interaction 
happens via the strong interaction.  Therefore, for midrapidity 
($|\eta|<$~0.35) hadronic or jet final states with small to intermediate 
energies, quark-gluon and gluon-gluon interactions are the dominant 
processes.  Consequently, longitudinal-double-spin asymmetries, 
$A_{LL}$, are sensitive to the gluon-spin contribution to the proton, 
$\Delta g(x,Q^2)$. The RHIC jet~\cite{Adamczyk:2014ozi} and neutral pion 
asymmetry measurements~\cite{Adare:2014hsq} at a 
center-of-mass energy, $\sqrt{s}$, of 200 GeV resulted in the first 
indication of a nonzero gluon-spin contribution to the nucleon spin when 
the jet and neutral-pion data was analyzed together with the DIS and 
semi-inclusive DIS results in a global 
analysis~\cite{deFlorian:2014yva,Nocera:2014gqa}. Subsequently, various 
measurements at a higher collision energy of 510 GeV have confirmed this 
nonzero gluon 
polarization~\cite{Adare:2015ozj,Adam:2018pns,Adam:2018cto,Adam:2019aml} 
and those combined with results at 
$\sqrt{s}=200$~GeV~\cite{Adamczyk:2016okk} have extended the parton 
momentum fraction $x$ coverage to lower values of approximately 
$10^{-3}$.

While the global fits clearly prefer a positive gluon polarization in 
the probed $x$ range, another direct experimental confirmation would be 
helpful. The addition of charged pion asymmetries with the help of 
different fragmentation of up and down quarks~\cite{deFlorian:2014xna} 
into $\pi^\pm$ provides this possibility. Because up and down quark 
polarizations are reasonably well known, the ordering of the positive, 
neutral and negative pion asymmetries immediately informs about the sign 
of the gluon spin. A positive gluon spin, coupled with the positive up 
quark polarization and negative down quark polarization would result in 
$\pi^+$ asymmetries to be the largest, followed by $\pi^0$ and, then, 
$\pi^-$. The charge-separated pion asymmetry results at 
$\sqrt{s}=200$~GeV have already been published~\cite{Adare:2014wht}.

In this paper, we report the charged pion longitudinal double spin 
asymmetries at $\sqrt{s}=510$~GeV that were extracted by the PHENIX 
experiment at midrapidity. The paper is organized as follows. In 
Sec.~\ref{experimentalsetup}, the PHENIX experiment and the detector 
components relevant for this result are described. In 
Sec.~\ref{analysisprocedure}, the analysis procedure for extracted 
charged pions and their double spin asymmetries at midrapidity is 
discussed. In Sec.~\ref{results}, the results are presented. Summary 
is given in Sec.~\ref{summary}.

\section{Experimental setup} \label{experimentalsetup}

In 2013, the PHENIX experiment at RHIC collected data from 
longitudinally polarized $p$$+$$p$ collisions at $\sqrt{s}=510$~GeV with 
an average polarization of 0.55 and 0.56 for the clockwise (blue)
and counter-clockwise (yellow) beams, respectively. 
An integrated luminosity of 108 $pb^{-1}$ was sampled for 
charged-pion asymmetry measurements at midrapidity.

The PHENIX detector is described in detail in 
Ref.~\cite{Adcox:2003zm}. Each of two nearly back-to-back arms of 
the central spectrometer covers a rapidity range $|\eta|<$~0.35 and 
an azimuthal range of $\Delta{\phi}=\frac{\pi}{2}$. The PHENIX 
detector elements used in this analysis include the drift chambers 
(DC), the pad chambers (PC), the ring imaging \v{C}erenkov (RICH) 
detector and the electromagnetic calorimeters (EMCal). The RICH, 
filled with CO$_2$ gas radiator, is used for charged-pion 
identification. The EMCal comprises two different types of 
calorimeters.  Six sectors are constructed with lead-scintillator 
(PbSc) towers in sampling configuration with depth of 0.85 
interaction lengths.  Two sectors are made of lead-glass towers with 
a depth of 1.05 nuclear interaction lengths.  Because the events 
sampled for this analysis are triggered via energy deposit 
thresholds, only the fraction of pions that shower in the EMCal are 
available.  Analysis is limited to the PbSc-triggered events, 
because the higher-energy thresholds result in lower background 
fractions than in the lead-glass towers.  Charged particle tracks 
are reconstructed with the DC and PC tracking system. These 
detectors also provide the momentum information of the tracks. A 
match between a projected track onto the EMCal and the location of 
deposited energy is required to veto charged tracks with 
mis-reconstructed momenta. The silicon-vertex detector surrounds 
the beam pipe with layers at nominal radii 2.6, 5.1, 11.8, 
16.7 cm with an acceptance of $|\eta|<$~1 and $\Delta{\phi}=0.8\pi$. 
The total material budget is 0.13 radiation lengths and the detector 
was not in operation in 2013.  This created a large source of 
electron background from conversions of direct and decay photons.

Additionally, two sets of 64 quartz-crystal radiators attached to 
photomultipliers located at $z$ positions of $\pm$144 cm and rapidities 
between 3.1 to 3.9 were used to trigger hard collision events and to 
select events within $\pm$30 cm of the collision vertex in the asymmetry 
analysis.  These beam-beam-counters and the zero-degree calorimeters 
were used together to evaluate the luminosities seen by the PHENIX 
detector. The zero-degree calorimeters comprise three sections of a 
hadronic calorimeter located at $\pm$18 m from the PHENIX interaction 
point are also used to monitor the polarization orientation and confirm 
that the polarization direction of the beams has been rotated to the 
longitudinal direction.

\section{Analysis procedure} \label{analysisprocedure}

\subsection{Data set and triggers}

\begin{figure}[thb]
\includegraphics[width=1.0\linewidth]{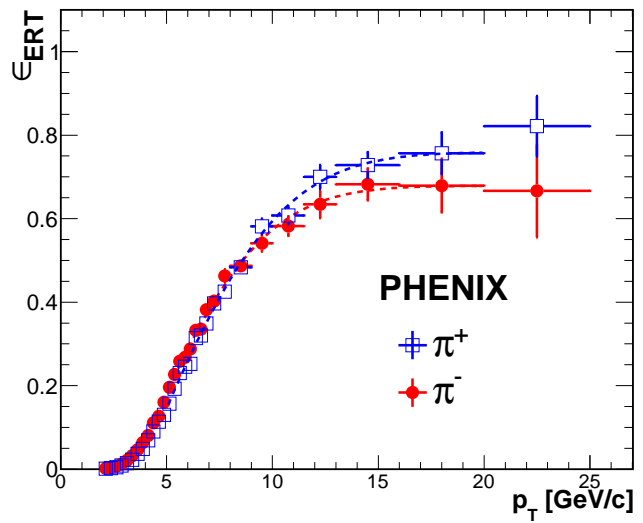}  
\caption{\label{fig:trigger}
Trigger efficiency curves of the EMCal-RICH trigger for positively 
charged (open [blue] squares) and negatively charged (closed [red] circles) 
pion candidates in the PbSc as a function of the transverse momentum of 
the track. The energy threshold of the trigger was at 3.7 GeV. Note that 
a cut on the ratio between cluster energy to reconstructed momentum 
($E/p$) was applied in pre-selection of the $\pi^{\pm}$ sample. The 
charge difference seen at higher $p_{T}$ originates from the momentum 
reconstruction which could not be perfectly calibrated in the high rate 
conditions of the 2013 data taking period.
}
\end{figure}

The 2013 detector configuration was similar to the published results at 
$\sqrt{s}=200$~GeV~\cite{Adare:2014wht} in 2009, except that the 
hadron-blind detector (HBD) was no longer installed. Due to the higher collision energy and collision rates in 2013, the 
energy thresholds of the EMCal triggers were increased by a factor of 
$\approx$2--3 compared to in 2009 and events were triggered by 
particles leaving at least 2.2, 3.7, 4.7 or 5.6 GeV energy deposits in 
the EMCal for the various trigger types. The lower energy threshold 
triggers were pre-scaled such that only a fraction of events satisfying 
the trigger requirements was recorded. An OR of all these triggers was 
used for the transverse momentum bins in the range 5 GeV/$c$ $<p_T<$ 
11~GeV/$c$, where the less pre-scaled higher threshold triggers are 
dominant. To minimize the background contribution for the highest 
transverse momentum bin (11 GeV/$c$ $< p_T <$ 15 GeV/$c$), the 2.2 GeV 
threshold trigger was not used. The trigger efficiency curves as a 
function of transverse momentum with energy threshold of 3.7 GeV for 
the PbSc are displayed in Fig.~\ref{fig:trigger} for $\pi^{\pm}$ 
candidates where also a preselection cut on the ratio between cluster energy 
to reconstructed momentum ($E/p$, to be described in detail below) was 
already applied. High $p_T$ charged pions punch through the EMCal with 
approximately a 50\% chance, depositing only a small fraction of their 
energy corresponding to the minimum-ionizing particles (MIPs) at 
$\approx$0.3 GeV due to their low probability of nuclear interactions 
in the detector. The pre-selection cuts for $\pi^{\pm}$ are blind to 
the MIPs interactions and consequently result in higher trigger 
efficiencies than for the case where all types of interactions are 
taken into account. Nonetheless, this analysis does not include MIPs, 
and the approach properly takes into account the $p_T$ dependence of 
trigger efficiency after applying pre-selection cuts.

\begin{figure}[thb]
\includegraphics[width=1.0\linewidth]{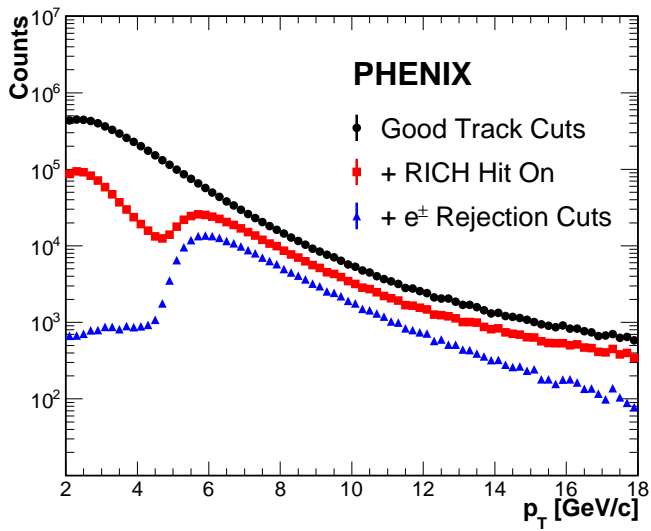}  
\caption{\label{fig:ptdist}
Pion candidate transverse momentum distributions after successively 
applying raw track criteria (closed [black] circles), RICH hit requirement 
(closed [red] squares) and electron rejection via $E/p$, matching and 
shower shape (closed [blue] triangles).
}
\end{figure}

\subsection{Charged pion identification and background estimation}
In addition to the trigger, a matching track in the drift chamber is 
required to be pointing to the EMCal tower that fired the trigger. The 
transverse momentum of the particle is determined by the bending of the 
track in the magnetic field before the DC. In addition, the 
reconstructed tracks are required to fire more than one photomultiplier 
by \v{C}erenkov light in the RICH. The threshold for pions is around 4.9 
GeV and until the kaon threshold of 17.3 GeV is reached the RICH fires 
only for pions and electrons (muons are not dominant and are already 
eliminated by the energy cut from the high energy threshold of trigger). 
To remove electrons as well as accidental track - EMCal cluster 
coincidences, the ratio between cluster energy and track momentum 
($E/p$) is required to be larger than 0.2 and smaller than 0.8, taking 
into account that most pions do not deposit all their energy in the 
electromagnetic calorimeter in contrast to electrons. 

\begin{figure}[thb]
   \includegraphics[width=1.0\linewidth]{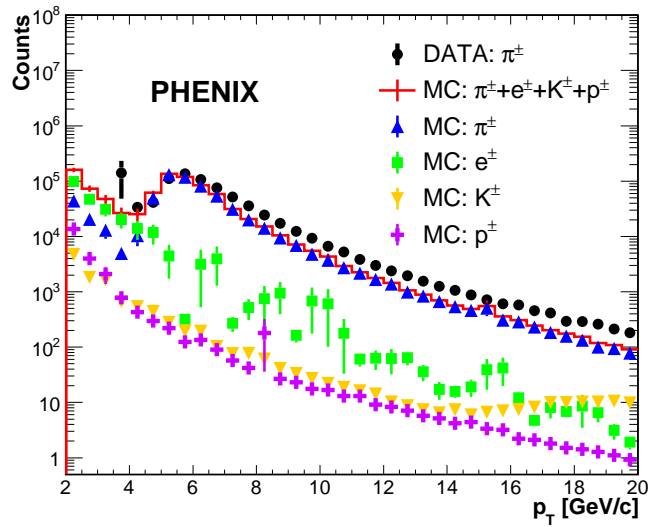}  
 \caption{\label{fig:pt2}
Comparison of reconstructed particle momentum distributions as a 
function of the transverse momentum in the data and MC simulations. The 
pion candidates of the data (closed [black] circles) are corrected for 
the trigger efficiency. The pion (closed [blue] triangles), electron 
(closed [green] squares), kaon (closed [yellow] inverted triangles), 
proton (closed [purple] crosses), and all (histogram [red] lines) 
contributions of the MC simulation are scaled by the luminosity for 
apple-to-apple comparison.
}
 \end{figure}

For the further rejection of electron background from the charged 
pion candidates, the probability that a cluster has developed via an 
electromagnetic shower processes (shower shape) was determined from 
fitting the well understood electromagnetic shower shape in the 
EMCal to the cluster in question. The shower shape probability was required to be less than 
0.1. The succession of the selection criteria on the raw charged 
particle spectra can be seen in Fig.~\ref{fig:ptdist}. A clear bump 
can be seen once the momentum is large enough for pions to emit 
\v{C}erenkov light. The contribution at momenta below the bump 
indicates remaining electrons and other accidental coincidences. 
After applying electron rejection cuts, their contributions are 
substantially reduced ($\approx$ 0.01--0.085). The remaining 
background in the higher transverse momentum range is studied with 
full MC simulations using {\sc pythia}\cite{Sjostrand:2006za} as 
event generator and {\sc geant3}~\cite{Brun:1994aa} for the detector 
description. Figure~\ref{fig:pt2} shows that at low transverse 
momenta below 5 GeV/$c$ the distribution is dominated by electrons, 
accidental pion coincidences, and (to a smaller extent) kaons and 
protons.  At higher transverse momenta, electrons are the dominant
background, which is small compared to pion signals until the RICH hit 
requirement becomes fulfilled by kaons as well.  The simulated 
contributions describe reasonably well both the signal-dominated 
region at higher transverse momenta and the background-dominated 
region below 5 GeV/$c$.

 \begin{figure}[htb]
 \includegraphics[width=1.0\linewidth]{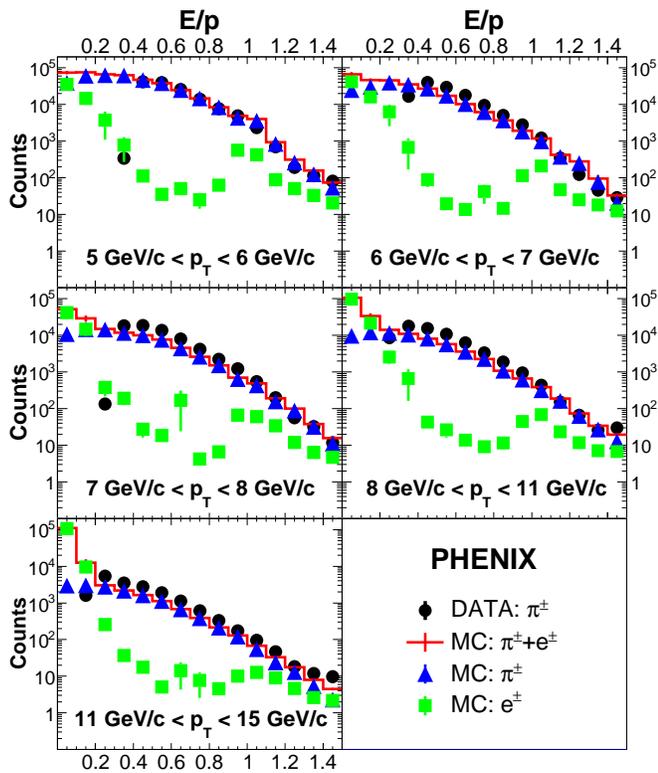}  
 \caption{\label{fig:ep}
Energy over momentum ratio for pion candidates in bins of transverse 
momentum. Reconstructed Data (closed [black] circles) are compared to 
luminosity scaled MC contributions by pions (closed [blue] triangles) 
and electrons (closed [green] squares) as well as their sum (histogram 
[red] lines). Due to the minimum energy requirement in the trigger, the 
data drops at very low values to zero. Note that there are huge tails 
from true electrons in lower $E/p$ regions. This is because electrons 
from photon conversion and/or decay-in-flight are reconstructed with 
higher transverse momentum and then the measured $E/p$ are lowered. 
These off-vertex electron backgrounds are eliminated using the $E/p$ 
cuts.
}
 \end{figure}

The relative size of pion signal and electron backgrounds is then 
further compared with data by studying the full $E/p$ range including 
the electron peak at ratios around unity where it is quite prominent. 
Based on this comparison, as seen in Fig.~\ref{fig:ep}, the nonpion 
background is found to be below a few percent. A Gaussian function for 
the electron peak and an Error function for the pion signal are fit to 
the $E/p$ distribution in each $p_T$ bin. The extracted parameter of the 
Gaussian was used to scale the electron background from the simulation. 
As the background level was found to be small, the scaling factor was 
varied by a factor of two in the background corrected asymmetries, 
variation was assigned as systematic uncertainty and the effect of the 
scale variation found to be small.

\subsection{Asymmetry analysis}

The selected pions are then separated by spin pattern, which determines 
whether the protons collided with the same or opposite helicities. These 
asymmetries are normalized for the fluctuations in luminosity from the 
bunch crossings with the same ($++$) helicity and opposite ($+-$) 
helicity, known as relative luminosity, 
$R=\mathcal{L}^{++}/\mathcal{L}^{+-}$ ($\approx$1.002):

\begin{equation}
A_{LL} = \frac{1}{P_BP_Y} \frac{N^{++} - R N^{+-}}{N^{++} + R N^{+-}}\quad,
\end{equation}

\noindent where $P_B$ and $P_Y$ are the average beam polarizations for the 
blue and yellow beam, respectively, and N is the 
number of charged pions from the bunch crossings with the same and 
opposite helicities. 

In 2013, nominal beam fills from injection to dump of beams at RHIC 
lasted eight hours. PHENIX DAQ system collected data in runs within the 
fill. Because the pre-scale of the trigger as well as the polarization 
values, which were calculated by the initial polarization and the rate 
of decrease of polarization as a function of time, changed on a 
run-by-run basis, the analysis is carried out separately for each run. 
The asymmetries are calculated for each run and each transverse momentum 
bin and are fit by a constant over all runs. During the 2013 RHIC running period the average beam polarizations $P_B$ and $P_Y$
were 0.55$\pm$0.02 and 0.56$\pm$0.02 for blue and yellow beams,
respectively~\cite{hjet}.

During the data-taking 16 different spin pattern combinations for the two beams were utilized 
to minimize systematic effects.  These several patterns were found 
to provide consistent asymmetries, based on T-tests between them, and 
therefore no systematic uncertainty was assigned due to the different 
patterns. 

To test for other potential systematic effects, the asymmetry 
calculation is repeated many times with randomized spin patterns for 
each run. The resulting asymmetry distributions for all iterations peak 
around zero with a Gaussian width given by the statistical uncertainties 
and the corresponding $\chi^2 /n.d.f.$ distributions of the fits center 
around unity.

Other systematic uncertainties include a global scale 
uncertainty of 6.5\% due to the accuracy of the beam polarization 
determination~\cite{hjet} and the transverse component of the beams, 
which has been found to be negligible for the double longitudinal spin 
asymmetries. The uncertainty on the asymmetries based on 
the relative luminosity extraction is ${\delta}A_{LL}=3.8{\times}10^{-4}$. 

The momentum scale uncertainty of the hadron transverse 
momentum has also been taken into account, but given the size of the 
transverse momentum bins used for the asymmetries, bin migration is 
minimal. The nonpion background has also been considered based on the 
background yields evaluated by comparing MC with data. The background 
asymmetry is estimated based on an electron enhanced data sample, which 
is found to be consistent with zero. The systematic uncertainty from the 
background asymmetry is evaluated by varying the background fraction 
after taking into account the evaluated background asymmetry mentioned 
above. These systematic uncertainties range from $2\times10^{-5}$ to 
$10^{-3}$.

\section{Results} \label{results}

\begin{figure}[th]
 \includegraphics[width=1.0\linewidth]{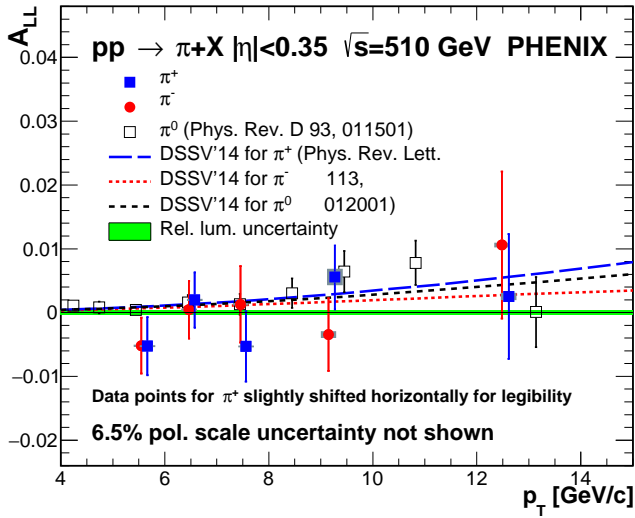}  
\caption{\label{fig:all}
Double-spin asymmetries $A_{LL}$ as a function of transverse momentum 
for positive (closed [blue] squares) and negative pions (closed [red] 
circles), as well as the previously 
published~\protect\cite{Adare:2015ozj} neutral pions (open [black] 
squares). The statistical uncertainties of asymmetries and the 
point-to-point systematic uncertainties from background are represented 
by the continuous lines and the gray bands, respectively.  The expected 
asymmetries based on the DSSV~\protect\cite{deFlorian:2014yva} fit 
(only from the 200 GeV data but none of the 510 GeV data) are displayed 
in the indicated line types.  The uncertainty bands on the fits are not 
shown as they affect all charges in similar ways.
}
\end{figure}

The resulting final double spin asymmetries are displayed in 
Fig.~\ref{fig:all} as a function of transverse momentum for positive and 
negative pions and compared to the previously published neutral pions. 
As can be seen, the results are consistent with the 
DSSV~\cite{deFlorian:2014yva} fit that has considered only the 200 GeV 
data but not the 510 GeV data. Due to the large statistical 
uncertainties, the sign of the gluon polarization in the probed $x$ 
region cannot directly be inferred from the ordering of the asymmetries 
for the three charges. However, it was found that the present results 
are consistent with the positive gluon polarization from the global 
fits. The reason for the comparatively low statistics for charged pions 
compared to neutral pions is the trigger requirement of having 
substantial energy deposited in the electromagnetic calorimeter, which 
happens only for a small fraction of charged pions. 

\begin{table}[ht]
\caption{Charged pion double spin asymmetries $A_{LL}$
in bins of transverse momentum $p_T$.
}
    \label{tab:allresults}
     \begin{ruledtabular} \begin{tabular}{cccccc}
$\pi^\pm$
& $p_T$ bin & $\langle p_T\rangle$ &  $A_{LL}$         & Stat. &  Syst.  \\
& [GeV/$c$] &    [GeV/$c$]         & [$\times10^{-3}$]
& [$\times10^{-3}$] & [$\times10^{-3}$] \\ \hline
$\pi^-$
& 5--6   &  5.55  &  -5.19 &  4.38 &  0.10 \\
& 6--7   &  6.47  &  0.45 &  4.51 &  0.10 \\
& 7--8   &  7.46  &  1.29 &  5.98 &  0.10 \\
& 8--11  &  9.15  &  -3.44 &  5.71 &  0.44 \\
& 11--15 &  12.48 &  10.60 &  11.52 &  0.30 \\
$\pi^+$
& 5--6   &  5.57 &  -5.26 &  4.57 &  0.08 \\
& 6--7   &  6.48 &  1.97 &  4.34 &  0.09 \\
& 7--8   &  7.46 &  -5.30 &  5.51 &  0.08 \\
& 8--11  &  9.17 &  5.58 &  5.08 &  1.20 \\
& 11--15 &  12.51 &  2.52 &  9.78 &  0.49 \\
    \end{tabular} \end{ruledtabular}
\end{table}

\begin{figure}[th]
\includegraphics[width=1.0\linewidth]{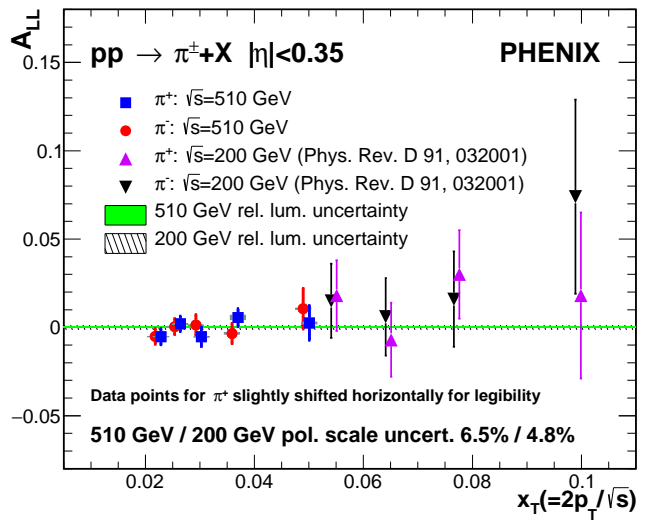}  
\caption{\label{fig:allxt}
Double spin asymmetries $A_{LL}$ as a function of $x_T=2p_T/\sqrt{s} $ 
for positive (closed [blue] squares) and negative pions (closed [red] circles) 
at $\sqrt{s} =$ 510 GeV as well as charged pions at 200 GeV (closed  
[purple] triangles and closed [black] inverted triangles). The data shown 
here are tabulated in~\protect\ref{tab:allresults}. 
}
 \end{figure}

In addition, one can also compare these data to the previously published 
measurements of charged pions at $\sqrt{s} = 200$~GeV. They are 
complementary because the hadrons detected at the same transverse 
momenta but at different center-of-mass energies probe different 
momentum fraction region. Therefore, the exact same measurement at 
higher collision energy of $\sqrt{s}=510$~GeV probes a lower value of 
$x$ than what was possible with the previously published data at 
$\sqrt{s} = 200$~GeV. While the experimentally measured transverse 
momentum contains a convolution of $x$ for both partons and the momentum 
fraction $z$ from the fragmentation process, the variable 
$x_T=2p_T/\sqrt{s} $ can act as a proxy for the $x$ ranges probed. 
Figure~\ref{fig:allxt} shows the measurements at 200 and 510 GeV and one 
can see the substantially lower $x_T$ reach. Based on {\sc 
pythia}~\cite{Sjostrand:2006za} simulations of charged pions in the 
rapidity range and transverse momentum ranges probed in this 
publication, mean $x$ values of $\approx$ 0.04--0.09 can be accessed. 
Despite the limited statistical precision, this additional information 
at lower $x$ will improve global fits of the gluon polarization when 
this data is included. The asymmetries are tabulated in Table 
\ref{tab:allresults}.

\section{Summary} \label{summary}

In summary, PHENIX has measured the charged pion double spin asymmetries 
at midrapidity ($|\eta|<$~0.35) in longitudinally polarized 
$p$$+$$p$ collisions at $\sqrt{s} = 510$~GeV. These measurements are 
sensitive to the gluon spin contribution to the total spin of the proton 
in $x$ range $\approx$ 0.04--0.09. The asymmetries are found to be 
consistent with global fits that have included only 200 GeV RHIC data, 
and a nonzero, positive gluon polarization in the $x$ region probed by 
RHIC has been found. In the proposed sPHENIX experiment~\cite{sphenix}, 
the hadronic calorimeter will greatly enhance triggering efficiency for 
charged hadrons and, therefore, significantly improve the statistical 
precision for charged pion measurements and make such direct evaluation 
of the gluon spin contribution possible.



\begin{acknowledgments}

We thank the staff of the Collider-Accelerator and Physics
Departments at Brookhaven National Laboratory and the staff of
the other PHENIX participating institutions for their vital
contributions.  We acknowledge support from the
Office of Nuclear Physics in the
Office of Science of the Department of Energy,
the National Science Foundation,
Abilene Christian University Research Council,
Research Foundation of SUNY, and
Dean of the College of Arts and Sciences, Vanderbilt University
(U.S.A),
Ministry of Education, Culture, Sports, Science, and Technology
and the Japan Society for the Promotion of Science (Japan),
Conselho Nacional de Desenvolvimento Cient\'{\i}fico e
Tecnol{\'o}gico and Funda\c c{\~a}o de Amparo {\`a} Pesquisa do
Estado de S{\~a}o Paulo (Brazil),
Natural Science Foundation of China (People's Republic of China),
Croatian Science Foundation and
Ministry of Science and Education (Croatia),
Ministry of Education, Youth and Sports (Czech Republic),
Centre National de la Recherche Scientifique, Commissariat
{\`a} l'{\'E}nergie Atomique, and Institut National de Physique
Nucl{\'e}aire et de Physique des Particules (France),
Bundesministerium f\"ur Bildung und Forschung, Deutscher Akademischer
Austausch Dienst, and Alexander von Humboldt Stiftung (Germany),
J. Bolyai Research Scholarship, EFOP, the New National Excellence
Program ({\'U}NKP), NKFIH, and OTKA (Hungary),
Department of Atomic Energy and Department of Science and Technology
(India),
Israel Science Foundation (Israel),
Basic Science Research and SRC(CENuM) Programs through NRF
funded by the Ministry of Education and the Ministry of
Science and ICT (Korea).
Physics Department, Lahore University of Management Sciences (Pakistan),
Ministry of Education and Science, Russian Academy of Sciences,
Federal Agency of Atomic Energy (Russia),
VR and Wallenberg Foundation (Sweden),
the U.S. Civilian Research and Development Foundation for the
Independent States of the Former Soviet Union,
the Hungarian American Enterprise Scholarship Fund,
the US-Hungarian Fulbright Foundation,
and the US-Israel Binational Science Foundation.

\end{acknowledgments}



%
 
\end{document}